# Sustaining dry surfaces under water

Paul R. Jones,[1] Xiuqing Hao,[1] Eduardo R. Cruz-Chu,[2] Konrad Rykaczewski,[3] Krishanu Nandy,[1]

Thomas M. Schutzius,[4] Kripa K. Varanasi,[5] Constantine M. Megaridis,[4] Jens H. Walther,[2,6]

Petros Koumoutsakos,[2] Horacio D. Espinosa,[1] Neelesh A. Patankar[1,*]

[1]Department of Mechanical Engineering, Northwestern University, Evanston, IL, USA.

[2]Institute of Computational Science, ETH Zürich, Zürich, Switzerland.

[3]School for Engineering of Matter, Transport and Energy, Arizona State University, Tempe, AZ, USA.

[4]Department of Mechanical and Industrial Engineering, University of Illinois at Chicago, Chicago, IL, USA.

[5]Department of Mechanical Engineering, MIT, Cambridge, MA, USA.

[6]Department of Mechanical Engineering, Tech. University of Denmark, Kgs. Lyngby, Denmark

[*]E-mail: n-patankar@northwestern.edu

**Rough surfaces immersed under water remain practically dry if the liquid-solid contact is on roughness peaks, while the roughness valleys are filled with gas. Mechanisms that**

**prevent water from invading the valleys are well studied. However, to remain practically dry under water, additional mechanisms need consideration. This is because trapped gas (e.g. air) in the roughness valleys can dissolve into the water pool, leading to invasion. Additionally, water vapor can also occupy the roughness valleys of immersed surfaces. If water vapor condenses, that too leads to invasion. These effects have not been investigated, and are critically important to maintain surfaces dry under water. In this work, we identify the critical roughness scale below which it is possible to sustain the vapor phase of water and/or trapped gases in roughness valleys – thus keeping the immersed surface dry. Theoretical predictions are consistent with molecular dynamics simulations and experiments.**

Superhydrophobicity occurs when surface roughness enhances non-wetting properties of hydrophobic solids.[1,2] Maintaining superhydrophobicity of rough textured surfaces has typically relied on the presence of trapped air pockets in the roughness valleys.[3] Keeping these surfaces practically dry (liquid minimally touching the solid surface) under water is challenging because the trapped air is found to deplete.[4-7] This depletion limits the utility of these surfaces in applications like drag reduction,[4,5,8] boiling,[9] among others. We investigate how immersed surfaces can remain practically dry. We postulate that it is essential to stabilize the vapor phase of water and sustain trapped gases in roughness valleys. There is a critical roughness scale, below which these mechanisms are effective. These are passive thermodynamic mechanisms that do not involve active generation[5] or exchange of gas.[10,11] We show that surfaces of hydrophobic solids retain non-wetting properties in the presence of sub-micrometer roughness. Theoretical predictions are consistent with molecular dynamics simulations, experiments, and observations of air-retaining insect surfaces.[12,13] We envision that this work will pave way to rationally design surface texture to manipulate the phase of one material adjacent to a surface – in this instance acquiring a vapor phase between a liquid and a textured solid surface, even when the liquid is not heated or boiled.

Although dry immersed rough surfaces are deemed viable, the underlying mechanisms that drive non-wetting to wetting transitions are not fully understood. Research on the well-known wetting behavior of non-immersed rough surfaces,[1,2] manifested in the form of liquid droplets beading up and moving with very little drag, has intensified in recent years.[3] In this case, the droplets reside on top of roughness peaks, while air occupies roughness valleys. This is the Cassie-Baxter (CB) state.[1] Maintaining the CB state will ensure a practically dry surface while immersed in a liquid. This is challenging,[4-7] as air in the roughness valleys can dissolve into the liquid if the liquid is undersaturated with air. Thus, in order to keep a surface practically dry under water, the gas phase in the roughness valleys must be sustained.

The thermodynamic analysis of underwater superhydrophobicity that accounts for only the surface energy has been theoretically studied.[14] To make robust surfaces that remain dry under water, the effect of sustaining vapor pockets also needs to be accounted in the thermodynamic analysis.[15,16] To elucidate the fundamental principles required to sustain gas pockets, we consider a typical cylindrical pore on a surface that is immersed under water (Figure 1A). When the surface is immersed under water, there will initially be air trapped in the pore (roughness valley). For this air to be sustained over a long period, it should be in chemical equilibrium with air dissolved in the ambient liquid. If the liquid is supersaturated with air, an air layer covering the surface may be achieved indefinitely.[10] However, if the liquid is undersaturated, then air within the pore will dissolve into the liquid.[17] Consequently, air pressure inside the pore will decline, and water will invade if the liquid-air interface cannot remain pinned at the top of the pore.[18-31] The invading liquid will lead to the wetting of the immersed surface.

Trapped air is not the only gas that can occupy the pore. At temperatures below the boiling point, the liquid phase is the lower energy state. However, a metastable vapor can evaporate from the meniscus (hanging at the top of the pore) and occupy the pore. This vapor inside the pore could eventually condense on the pore walls, thus providing another pathway, via condensation, to wet the pore. Will the metastable vapor occupy the pore and keep it dry or will it condense in the pore to make it wet? This is a critical consideration, hitherto unresolved, and which is essential to enabling practically dry surfaces immersed in undersaturated liquids.

We term the phenomenon of sustaining the metastable vapor in the pore as vapor-stabilization. This is important because it permits sustaining the vapor phase without actually having to boil the liquid. This mechanism has been considered to stabilize the film-boiling mode even at low superheats.[32] Analysis of the energetics of the competing scenarios (wetting vs. non-wetting) yields to the following condition to avoid liquid invasion and keep the pore dry:[15, 16]

$$D < -\frac{4\sigma_{lg}}{p_l - p_g}\cos\theta_e \quad \text{OR} \quad p_l < p_g - \frac{4\sigma_{lg}}{D}\cos\theta_e, \qquad (1)$$

where $D$ is the pore diameter, $p_l$ the liquid pressure, $p_g$ the pressure of the gas in the pore, $\sigma_{lg}$ the liquid-gas surface energy, and $\theta_e$ the equilibrium contact angle of a liquid drop on a flat solid surface of the same material as the pore. Typically, the gas will be a combination of trapped air and the vapor phase of the liquid, both of which should be in chemical equilibrium with the dissolved air in the liquid and the liquid itself, respectively (see Supplementary Figures S1-S2).[15, 16] Here, the pore is assumed to be deep enough so that the curved liquid-gas interface, hanging at the top of the pore, does not touch the bottom.[21] Equation (1) shows that for a given liquid pressure, the pore diameter should be smaller than a critical value to keep the liquid out of the pore, maintaining it dry. For example, assume that all air has dissolved out of the pore due to undersaturation of the liquid and the only gas in the pore is the vapor phase in chemical equilibrium with liquid water at room temperature and standard atmospheric pressure. This represents a vapor-stabilized scenario that would keep the pore dry. In this case, $p_l$ = 101.325 kPa, $p_g \approx$ 3.17 kPa,[15, 16] $\sigma_{lg}$ = 71.7 mN/m, and $\theta_e = 110^0$ (typical value attained by hydrophobic chemical coatings) yield a critical pore diameter of 1 μm. Equation (1) can also be used to predict the liquid pressure, above which the vapor will not be stabilized and liquid invasion will occur. This is plotted in Figure 1B for cases with $D$ = 10 nm.

Based on the above analysis, we predict that practically dry rough surfaces are possible in water, even after trapped air has fully depleted, due to the stabilized vapor phase of the liquid in the roughness valley. We estimate that, for typical liquid pressures, this will be feasible for pore diameters (roughness spacing) that are hundreds of nanometers or less, but not for roughness scales of tens of microns or larger. These conclusions based on pore-type geometries can be extended to pillar-type geometries without fundamental difficulty.[15, 16] In the remaining sections, we verify the above predictions using molecular dynamics simulations, experiments, and observations of air-retaining insects.

Molecular dynamics (MD) simulations using NAMD[33] 2.9 software were used to verify the liquid invasion pressures predicted using equation (1). To simulate an immersed rough surface, a 10 nm diameter cylindrical pore is assembled using VMD[34] software, with periodic boundary conditions for the overall domain (Figure 1A). The pore is solvated with SPC/E[35] water molecules residing initially outside the pore (on top of the roughness peaks). A rigid surface (piston) is used to apply pressure to the liquid water pooled above the pore. The pore assembly and meniscus trajectories are shown in Figure 1 (also see Supplementary Figures S8-S9). The MD results for invasion pressures applied at the piston, compared with theoretical predictions from equation (1), are shown in Figure 1B for temperatures of 300 K, 375 K, 450 K, and 501 K. For each temperature (from low to high), the corresponding pressures: $p_l$ = 107.79 bar, 88.19 bar, 73.49 bar, and 68.59 bar, respectively, demonstrate a resistance to liquid invasion, and hence, an immersed surface that remains practically dry. At the same respective temperatures but higher applied pressures, liquid invades the pore. A temperature of 501 K is used to allow a significant amount of vapor to accrue within the pore. A contact angle of 119.4° (accurate to within 9.07°) is determined from the angle between the meniscus and vertical pore walls at a temperature of 300 K (see Supplementary Section 4 for details). To demonstrate robustness against liquid invasion into the pore, we additionally simulate pores that are initially half-filled with water. The non-wetting behavior of these pores is consistent with simulations of water initially outside the pore. This is shown in Figure 1C for two simulations at a temperature of 501 K and 68.59 bar liquid pressure. We also note that at this temperature and pressure, the liquid is below its boiling point; yet the pore becomes occupied by the metastable vapor, as predicted. This method of using texture to control phase may potentially be extended to other phase transformations of water as well. In fact, molecular dynamics simulations indicate that condensation, or full wetting, can be achieved using rough hydrophilic surfaces at conditions above the boiling point of water (see Supplementary Figure S10). At these conditions, the presence of vapor is expected due to boiling, contrary to our results.

Physical experiments are conducted to establish the viability of keeping immersed surfaces dry. Scanning electron microscope (SEM) images of the samples before immersion are shown in Figure 2. Results for each experiment are reported in Table 1 and summarized below. Details are provided in Supplementary Sections 1-3, Supplementary Table S1, and Supplementary Figures S3-S7.

Aging experiments: Samples are immersed in a beaker of deionized water and shielded from external debris by covering the beaker top. Small holes are made in the cover to keep the system open to the environment. The optical property of total internal reflection is used to distinguish a state where there is a significant gas phase between the liquid and the solid surface. Samples are then removed from the beaker and tested for hydrophobic retention via water droplets. Surfaces that remained practically dry under water did not retain any water film when the samples were removed from water. Figure 3A shows an anodized aluminum substrate coated with a polymeric composite coating consisting of PVDF/PMMA (polymer matrix) and silica nanoparticles (filler).

Degassing in vacuum desiccator and vacuum oven: Samples are immersed in a beaker of water, and then placed into a vacuum desiccator to remove dissolved air. Samples are additionally degassed in a vacuum oven using a similar procedure. The degassing process for the zinc oxide coated sample is shown in Figure 3B. If surfaces maintain total reflection sheen and come out dry, we conclude that wetting was prevented even after air was depleted from the roughness valleys. This implies that the liquid does not condense in the roughness valleys; instead, the valleys remain dry with possibly the vapor phase in it.

Imaging the water-solid interface: Direct cross-sectional imaging of water-solid interfaces using cryostabilization, in combination with, cryogenic Focus Ion Beam milling and SEM imaging was recently demonstrated for liquid droplets.[36] In the present work, we adapted the same technique to image water-solid interfaces of superhydrophobic surfaces submerged below a few millimeters of degassed water. Images of the frozen water-solid substrate interface and its dependence on surface roughness spacing is shown in Figure 4. Liquid invasion was observed for micron scale roughness spacing, whereas, no invasion was observed for nanometer scale roughness spacing, as predicted.

In each experiment, immersed surfaces with hundreds of nanometer or less spacing remained practically dry. Samples with micron-size feature spacing became wet. The only discrepancy comes from the aluminum substrate sample, where wetting was observed in the vacuum oven after five days. We attribute this to multiple particle sizes defining the surface geometry (see Supplementary Section 1.1 for sample details). The experimental results presented in this work are at standard pressure conditions. Future interrogations should investigate the effect of pressurization, shear, and other factors on the wettability of the surface.

In addition to our experiments, others have observed consistent results in air retaining insect surfaces. Balmert et al.[12] conducted immersion experiments with air-retaining insect surfaces. Surface roughness on these insects is a result of hair spacing. Insect surfaces that remained dry the longest all had hair spacing of hundreds of nanometers or less (see Supplementary Table S2), as predicted here.

Observations of air-retaining insect surfaces, experiments with fabricated surfaces, and molecular dynamics simulations have all shown support for our proposition that sub-micron or smaller scale roughness is essential to maintaining dry surfaces under water. Small length scale roughness is necessary for stabilizing the vapor phase of water, and may serve as precedence for achieving general phase control of fluids using rough surfaces.

**Methods Summary**

**Molecular dynamics simulation.** The simulation consists of 301,228 atoms, 256,857 of which are water. The pore (Figure 1A) consists of two flat parallel graphene sheets and a carbon nanotube. The Extended Simple Point Charge (SPC/E[35]) water model is used with SETTLE[37] for rigid bonds. The Lennard Jones (LJ) carbon-carbon interactions are $\varepsilon_{CC}$ = -0.0565 kcal mol$^{-1}$ and $\sigma_{CC}$ = 3.23895 Å. The piston LJ interactions are $\varepsilon_{piston}$ = -0.1291 kcal mol$^{-1}$ and $\sigma_{piston}$ = 3.23895 Å. Carbon hydrophobicity is tuned using the oxygen-carbon LJ well-depth[38], i.e. $\varepsilon_{OC}$ = -0.0599 kcal mol$^{-1}$ for hydrophobic surfaces and $\varepsilon_{OC}$ = -0.1205 kcal mol$^{-1}$ for hydrophilic surfaces. Note: LJ well-depths in NAMD are negative by convention. Remaining non-bonded cross-interactions are defined by the Lorentz-Berthelot mixing rules. A cutoff radius of 12.0 Å and switch distance of 10.0 Å is used for all non-bonded interactions. The Particle Mesh Ewald algorithm calculated full electrostatic interactions every time step. A constant temperature is maintained using a Langevin thermostat[39] with a damping coefficient of 0.01 ps$^{-1}$. Carbon surface atoms are fixed, and piston atoms are constrained with a harmonic spring in the *x-y* plane using a force constant of 10 kcal mol$^{-1}$. Water within the nanopore is thermally equilibrated for at least 5 ns, with no applied pressure. For the half-filled nanopore simulations, atom velocities are reassigned during the initial configuration. This is done to prevent full wetting due to inertia from a prior state. The contact angle is measured in accordance with Ref [40] using bin sizes of 3.5533 Å fitted with a third order polynomial over 462 frames (924 picoseconds). See Supplementary Section 4 for details.

**Material fabrication.** Fabrication procedures for each material sample can be found in Supplementary Section 1.

**Degassing experiments.** The vacuum desiccator (420220000 Space Saver Vacuum Desiccator 190 mm Clear) reached a target pressure of 21.33-26.34 kPa during the day. The vacuum pump ran intermittently for 5-10 minutes, and then turned off for three hours. This occurred throughout the workday. At night, the pump is turned off while the vacuum desiccator remained closed. The chamber pressure increased overnight to 47.37 kPa the next morning due to leakage. Samples are additionally degassed in a vacuum oven (Model 281A Isotemp Vacuum Oven by Fisher Scientific) using a similar procedure as the vacuum desiccator. The pressure of the vacuum oven is kept at 2.0 kPa, which is below the boiling point of water. Samples are left in the closed oven over night and further degassed the following day.

Table 1. Experiment results of immersed surfaces.

| Material | Structure spacing | Liquid pressure | Observation: Dry/Wet (duration of experiment) |
|---|---|---|---|
| *Aging Experiments* | | | |
| Aluminum substrate[41] | 10 nm | Ambient | Dry (127 days) |
| Aluminum substrate[41] | 10 nm | Ambient | Dry (50 days) |
| Aluminum substrate[41] | 250 nm | Ambient | Wet |
| *Degassing in Vacuum Desiccator* | | | |
| Aluminum substrate[41] | 250 nm | All samples: 21.33-26.34 kPa (day), 47.37 kPa (night) | Wet (30 hours) |
| Zinc oxide | 40-80 nm | | Dry (~3 days) |
| Zinc oxide | 100-150 nm | | Dry (~3 days) |
| Silicon nanograss[42] | 100 nm | | Dry (5 days) |
| Silicon microposts[42] (Test #1, #2) | 5 μm | | Dry (5 days), Wet (3 days) |
| Silicon microposts[42] (Test #1, #2) | 25 μm | | Dry (5 days), Wet (3 days) |
| Silicon microgrooves (Two samples) | 3 μm, 12 μm | | Wet (3 days) |
| *Degassing in Vacuum Oven* | | | |
| Aluminum substrate[41] (Test #1, #2) | 10 nm | 2.0 kPa | Dry (4 hours), Wet (5 days) |
| Zinc oxide | 40-80 nm | 2.0 kPa | Dry (1.5 hours) |
| Zinc oxide | 100-150 nm | 2.0 kPa | Dry (1.5 hours) |
| Silicon nanograss[42] | 100 nm | 2.0 kPa | Dry |
| Silicon microposts[42] (Two samples) | 5 μm, 25 μm | 2.0 kPa | Wet |
| Silicon microgrooves (Two samples) | 3 μm, 12 μm | 2.0 kPa | Wet |
| *Imaging the water-solid interface* | | | |
| Silicon nanowire forest[36, 43] | 100-400 nm | A few Torr | Dry |
| Silicon microposts[42] (Two samples) | 5 μm, 25 μm | A few Torr | Wet |

Aging, degassing, and imaging experiments were conducted on various samples of rough hydrophobic solids immersed in water. Observations for each sample reflect the state of the surface at the conclusion of the experiment, even if non-wetting behavior is initially exhibited. Surfaces with sub-micron or less spacing tended to remain dry, whereas, surfaces with micron spacing became wet, as predicted.

**Supplementary Information Available.** Details of material fabrication, experimental methods, and simulation may be found in the supplementary information.

**Acknowledgments.** N.A.P, H.D.E, and K.N. acknowledge support from the Initiative for Sustainability and Energy at Northwestern (ISEN). X.H. acknowledges support from the Chinese Research Council. N.A.P and P.R.J. acknowledge computing resources from Northwestern University's high performance computing system (QUEST). K.R. acknowledges NIST for access to electron microscopy resources, and Dr. Albert Davydov and Dr. Sergiy Krylyuk from NIST for providing the VLS silicon nanowire samples.

**Author Contributions.** N.A.P. conceived the research. N.A.P, C.M.M., J.H.W., and P.K. planned the research. P.R.J. and E.R.C.C. performed simulations. P.K., J.H.W., and N.A.P. led the computational effort. T.M.S., X.H., and K.N. fabricated samples. K.K.V. and C.M.M. led the fabrication effort. X.H. and K.N. carried out the immersion experiments. N.A.P and H.D.E led the immersion experiments. K.R. did microscopy experiments. K.R. and K.K.V. led microscopy experiments. N.A.P. led the theoretical and the overall research effort. All authors contributed with discussions, analyses, and the interpretation of results. P.R.J. and N.A.P. wrote the manuscript. All authors edited the manuscript.

**Author Information.** Correspondence and requests for materials should be addressed to N.A.P. (n-patankar@northwestern.edu).

**Competing Financial Interests:** The authors have no competing interests or other interests that might be perceived to influence the results and/or discussion reported in this paper.

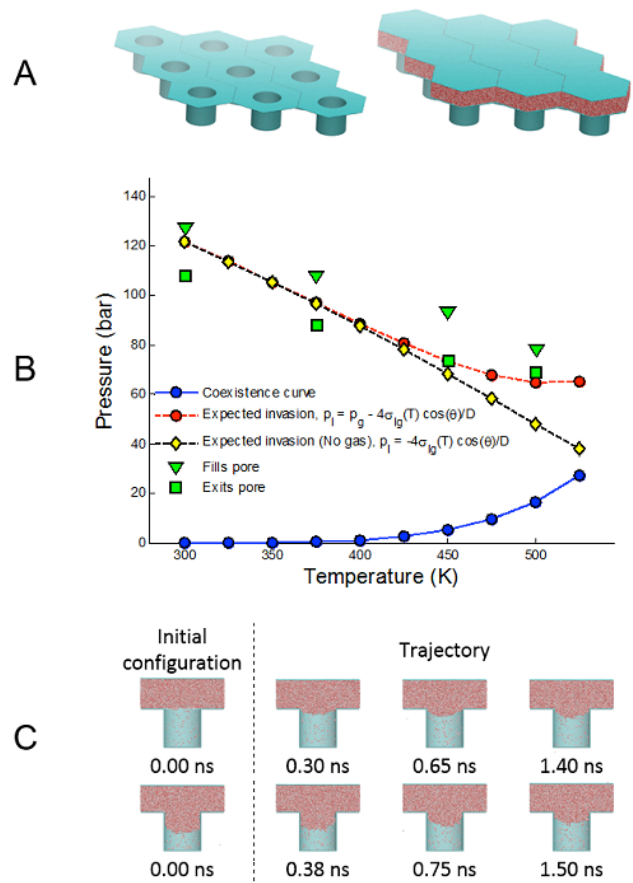

**Figure 1.** (A) Molecular dynamics model of a cylindrical pore surface with periodic boundary conditions. Water is placed on top of the textured surface. A rigid surface (piston) is used to apply pressure to the liquid water. (B) Liquid-vapor phase diagram for pore simulations. Stabilization and invasion pressures applied by the piston for an initially unfilled and initially half-filled pore were the same. The coexistence curve of the SPC/E water model obtained from the publicly available NIST Standard Reference Simulation Website[44] is shown. Expected liquid invasion pressures were determined by equation (1) using a calculated liquid-solid contact angle of 119.4° (accurate to within 9.07°), with surface tensions obtained from Sakamaki et al.[45] (C) Molecular dynamics simulations of a hydrophobic pore demonstrating non-wetting at 501 K and 68.59 bar applied pressure. The top row simulation begins with an empty pore; the bottom row simulation begins with a half-filled pore.

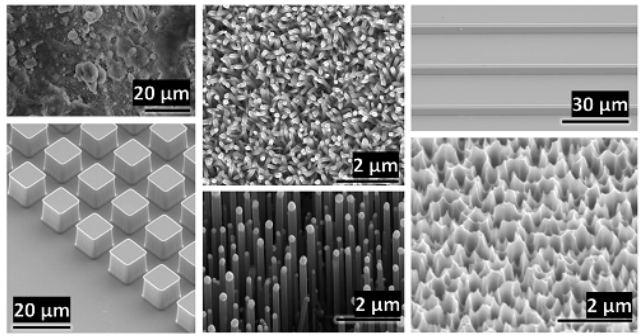

**Figure 2.** SEM images of the material surfaces used in our experiments. Left column: (top) polymer/nanoparticle composite coating on aluminum substrate, (bottom) silicon square microposts. Middle column: (top) zinc oxide film on silicon substrate, (bottom) silicon nanowire forest. Right column: (top) silicon microgrooves, (bottom) silicon nanograss.

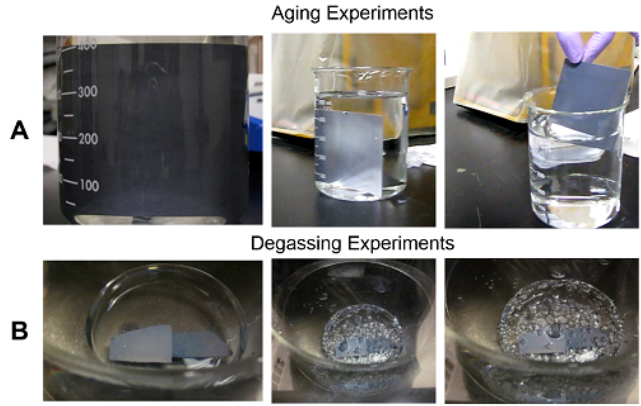

**Figure 3.** Experiments used to validate non-wetting behavior under water. (A) Anodized aluminum substrate coated with PVDF/PMMA and silica nanoparticles (10 nm structure spacing) after 127 days under water. The left image was taken orthogonally to the sample surface; the middle image is a side view that reveals a sheen caused by the thin gas layer between the surface and the water; the right image shows a dry sample upon retrieval from the bath. (B) Process of degassing air from the zinc oxide coated sample.

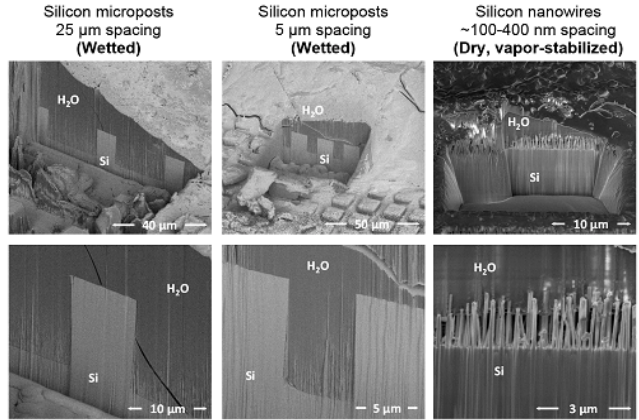

**Figure 4.** Direct nanoscale imaging of water-solid interfaces. Left: Wetted surface with 25 μm pillar spacing. Middle: Wetted surface with 5 μm pillar spacing. Right: Dry surface with submicron pillar spacing. Abbreviations: Frozen water ($H_2O$), Silicon substrate (Si).

# Supplementary Information

# Sustaining dry surfaces under water

Paul R. Jones,[1] Xiuqing Hao,[1] Eduardo R. Cruz-Chu,[2] Konrad Rykaczewski,[3] Krishanu Nandy,[1] Thomas M. Schutzius,[4] Kripa K. Varanasi,[5] Constantine M. Megaridis,[4] Jens H. Walther,[2,6] Petros Koumoutsakos,[2] Horacio D. Espinosa,[1] Neelesh A. Patankar[1,*]

[1]*Department of Mechanical Engineering, Northwestern University, Evanston, IL, USA.*

[2]*Institute of Computational Science, ETH Zürich, Zürich, Switzerland.*

[3]*School for Engineering of Matter, Transport and Energy, Arizona State University, Tempe, AZ, USA.*

[4]*Department of Mechanical and Industrial Engineering, University of Illinois at Chicago, Chicago, IL, USA.*

[5]*Department of Mechanical Engineering, MIT, Cambridge, MA, USA.*

[6]*Department of Mechanical Engineering, Tech. University of Denmark, Kgs. Lyngby, Denmark*

[*]*E-mail*: n-patankar@northwestern.edu

# Contents

## List of Supplementary Tables and Legends

## List of Supplementary Figures and Labels

# 1 Material Fabrication

Coated aluminum substrate samples were fabricated by the University of Illinois at Chicago (UIC). Zinc oxide coated silicon samples were fabricated by Xi'an Jiaotong University. Silicon nanowire forest, Silicon square microposts array, and silicon microgrooves samples were fabricated by the Massachusetts Institute of Technology (MIT) and the National Institute of Standards and Technology (NIST).

## 1.1 Polymer-nanoparticle composite coatings on aluminum

Materials: Acetone (ACS reagent, ≥ 99.5%), *N*-methyl-2-pyrrolidone (NMP, 99.5 wt. %), silicon dioxide nanopowder (99.5% metals basis, 5-15 nm), poly (methyl methacrylate) powder (PMMA, ~996,000 Da), and poly (tetrafluoroethylene) powder (PTFE, < 1 μm) were all obtained from Sigma-Aldrich. Poly (vinylidene fluoride) powder (PVDF, melt viscosity: 2,350-2,950 Pa s, melting point: 155-160 ºC) was obtained from Alfa-Aesar. Hydrophobic fumed silica (HFS, specific surface area (BET): 170±20 $m^2\ g^{-1}$; Aerosil® R 9200) was obtained from Evonik Industries. Previous analysis showed the primary feature size of PTFE particles to be 260±54 nm[1].

Fabrication: Separately, 10 wt. % stock solutions of PVDF and PMMA were generated by dissolving PVDF and PMMA powder in NMP and acetone, respectively, under slow mechanical mixing at room temperature overnight. In a typical case, a PTFE or HFS suspension was formed by combining 0.75 g of particles and 11 g of acetone in a 20 mL vial; this entire suspension was then bath sonicated (Cole-Parmer, 70 W, 42 kHz) for several minutes. Once a stable suspension was formed, 0.62 g of 10 wt. % PVDF and 0.62 g of 10 wt. % PMMA were added to it, and the entire dispersion was shaken mechanically at room temperature. The dispersion was then spray deposited with an airbrush (Paasche VL siphon feed, 0.73 mm spray nozzle) onto aluminum plates (i.e., mirror finished anodized aluminum) using compressed air (2.1 bar) at a fixed distance of ~10 cm. Initially the coatings were dried with a heat gun (Proheat® Varitemp® PH-1200, 1300 W max) for a minute and were then placed in an oven at 70 ºC for 60 min to completely dry, thus forming a superhydrophobic, self-cleaning coating. See Supplementary Table S1 for a description of the individual ingredients and their concentration in the dispersions used for spray. The previously mentioned procedure refers to coatings NC1 (HFS containing) and NC2 (PTFE containing).

## 1.2 Zinc oxide coated samples

Hydrothermal growth: First, a zinc oxide (ZnO) seed crystal layer of about 30~50 nm is prepared on the silicon surface by radio frequency (RF) magnetron sputtering system (explorer 14, Denton Vacuum, USA). The parameters of the sputter are as follows: power 150 W, flowrate Ar 20 sccm, and sputtering time 6 min. Second, ZnO nanorod arrays are synthesized on a seed crystal layer by hydrothermal method. The aqueous solution is a mixture of zinc nitrate hexahydrate ($Zn(NO_3)_2 \cdot 6H_2O$, 25 mM or 50 mM), hexamethylenetetramine (HMT, $C_6H_{12}N_4$, $Zn^{2+}$ and HMT are kept at the same 1:1 molar ratio), and deionized water (100 mL). ZnO nanorods were then synthesized at a temperature of 95°C for 1-3 hours.

Surface treatment: A 1% solution of F8261 (1H,1H,2H,2H-perfluoroalkyltriethoxysilanes) and methanol were used to chemically treat the surface. The surface was immersed in the solution for 26 hours, and subsequently heated for 1.5 hours at a temperature of 150 °C. Roll-off tests show the two samples are superhydrophobic, as indicated by water droplets easily rolling off the sample surface.

### 1.3 Silicon nanowire forest

Silicon nanowires were grown in a custom-designed horizontal hot-walled chemical vapor deposition reactor at 850°C using a $SiCl_4/H_2/N_2$ gaseous mixture. Gold nanoparticles formed on a Si(111) substrate by annealing a 5 nm thick gold film were used to catalyze the vapor-liquid-solid growth. To produce the superhydrophobic surface, grown nanowires with height of about 2.5 μm and diameters between 50 nm and 200 nm were modified using vacuum deposition of 1H,1H,2H,2H-Perfluorodecyltrichlorosilane (Alfa Aesar). Further details of the nanowire growth and surface modification can be found elsewhere.[2,3]

### 1.4 Silicon square micropost array

350 μm thick Si n-(100) substrates were patterned via photolithography and etched via Deep Reactive Ion Etching to obtain arrays of 10 μm tall square microposts with widths of 10 μm and edge-to-edge spacing of 5 μm and 25 μm. After cleaning the samples with Piranha solution, the samples were coated with octadecyltrichlorosilane (Sigma Aldrich) using a solution deposition method, rendering the surface hydrophobic.[4]

### 1.5 Silicon microgrooves

Two silicon microgroove samples were fabricated using the same procedure as the silicon square micropost array. The groove thickness, height, and spacing was 3 μm, 5 μm, 3 μm for the first sample and 3 μm, 5 μm, 12 μm for the second sample.

### 1.6  Silicon nanograss

Fabrication for the silicon nanograss[4] samples can be found elsewhere.

## 2  Experimental Methods

### 2.1  Aging experiments

The two aluminum substrate surfaces appeared matte black, when viewed from the normal direction, at all times in the experiment. When initially submerged, both substrates appeared to have a trapped layer of gas over their surface. The layer of gas in the 250 nm spacing surface appeared significantly reduced within a few hours. After seventy-two hours, no presence of the trapped gas could be observed. In comparison, the 10 nm surface showed almost no visual change in the quality of the trapped gas layer over a comparable period. The 10 nm spacing surface maintained the gas layer for 127 days (termination of experiment). The sample was then removed from the beaker and visually inspected. The majority of the sample remained dry, despite having been immersed in water for over four months. This was demonstrated using a pipette to place drops of water on the surface held at an angle; the droplets rolled off the surface upon contact

### 2.2  Repeated aging experiment of 10 nm spacing sample

The aging experiment with 10 nm spacing on the aluminum substrate was repeated for 50 days using a different sample with the same structure spacing provided by UIC. The results were consistent with the previous samples subjected to water exposure for 120 days.

### 2.3 Degassing experiments

The aluminum substrate sample with 250 nm structure spacing was subjected to five rounds of degassing in the vacuum desiccator. The aluminum substrate sample with 10 nm structure spacing was degassed twice over a period of 4 hours in the vacuum oven. The silicon nanograss sample was degassed twice in the vacuum oven.

## 3 Direct Nanoscale Imaging of Water-Solid Interfaces

See Supplementary Figure S7 for a schematic of the imaging procedure. The direct interface imaging experiments were carried out using an FEI Nova Nanolab 600 Dual Beam equipped with a Quorum PP2000T cryo-transfer system. About 7 mm by 7 mm pieces of the superhydrophobic surfaces were attached using double stick copper tape to a 10 mm diameter copper stub. About 5 mm tall piece of 9.5 mm (3/8 inch) inner diameter vinyl tubing was fitted around the copper stub. This tubing provided a watertight seal around the copper stub, creating a temporary water container that protruded about 3 mm above the sample. This temporary sample holder was submerged in a beaker filled with 30 mL of distilled water, which was placed in the vacuum desiccator used to make liquid nitrogen slush. The pressure within the chamber was reduced using a roughing pump until intensive bubble formation was observed. To prevent the water from freezing, bubbling intensity was reduced by leaking in nitrogen gas. After one minute of degassing, the vacuum desiccator was vented and the sample holder was carefully removed from the water beaker. The 3 mm water layer above the sample was subsequently frozen by placing the bottom part of the copper stub in contact with liquid nitrogen. We note that in our experiments the freezing rate was slower than during liquid nitrogen plunge freezing. Thus, some deformation of the liquid-gas-solid interface is expected due to crystallization.

The copper stub with ice-covered sample was attached to cryo-FIB/SEM shuttle and plunge frozen in liquid nitrogen slush. This proved to be necessary to prevent sticking of the transfer rod to the shuttle due to condensate freezing around the rod's thread. Next, the frozen sample was moved into the transfer chamber. The chamber was pre-evacuated to a pressure of $10^{-3}$ Pa and pre-cooled to a temperature of -180°C. To prepare the sample for FIB/SEM imaging, the thick ice layer was mechanically broken up using a metal cryo-manipulator and partially sublimated by raising the sample temperature to -90°C for a period of twenty to thirty minutes. To prevent charging of the sample, in situ 20 nm to 30 nm platinum deposition was achieved by plasma sputtering in Argon gas environment with 8 mA current for two minutes. After transfer to the cryo-FIB/SEM chamber, the samples were exposed to a 10 s pulse of $C_9H_{16}Pt$ gas precursor at 28°C. The resulting hundreds of nanometers to a few micrometers thick coating was cured by exposure to ion beam with energy and current of 30 keV and 2.6 nA, respectively. This additional coating (whitish-grey in color, covering the outermost surface shown in Figure 4) aided uniform milling during the cross-sectioning step.

FIB milling was performed with sample tilt of 52°, ion beam energy of 30 keV, an ion current of 0.093 nA to 21 nA, and a dwell time of 1 µs per pixel. The structure cross-sections were obtained by first FIB milling deep trenches at an ion beam current of 21 nA. Subsequently the surfaces were polished at 0.48 nA to 2.6 nA. All cross-sections were imaged at 52° tilt in the objective lens immersion mode using backscattered electrons with electron beam energy and current of 1 keV and 0.46 nA, respectively.

# 4  Measuring Contact Angles in Molecular Dynamics Simulations

A single contact angle, obtained from the vapor-stabilization case at 300 K, 107.79 bar pressure is used to estimate the critical contact angle used in Eqn. 1. Oxygen atoms at the liquid-vapor interface (2 Angstrom thick) were superimposed onto a single plot over the last 924 picoseconds of simulation. The interfacial oxygen were discretized in accordance with Ref[5] using bin sizes of 3.5533 Å. The contact angle was calculated by fitting a third order polynomial to the average radius within each bin. Using this method, the local contact angle was 119.4°.

To estimate the error associated with this measurement we rearrange Eqn. 1 in terms of $\sigma_{lg}$. The surface tension, $\sigma_{lg} = -(p_l-p_g)D(4\cos\theta)^{-1} = -(\Delta p)D(4\cos\theta)^{-1}$. At the same temperature we expect $\sigma_{lg}$ to remain constant, thus, $\sigma_{lg} = -(\Delta p)_{Out}D(4\cos\theta_{Out})^{-1} = -(\Delta p)_{In}D(4\cos\theta_{In})^{-1}$ where $(\Delta p)_{Out} = p_l-p_g$ and $p_l$ is the pressure for vapor-stabilization. $(\Delta p)_{In} = p_l-p_g$ where $p_l$ is the pressure for liquid invasion. The contact angles, $\theta_{Out}$ and $\theta_{In}$ correspond with the vapor-stabilization pressure and liquid invasion pressure respectively. Approximating the contact angle at the point of liquid invasion gives $\theta_{In} = \cos^{-1}[((\Delta p)_{In}/(\Delta p)_{Out})\cos\theta_{Out}]$. Using vapor-stabilization and liquid invasion pressures at each isotherm yields $\theta_{In}$ = 125.51°, 126.92°, 128.51°, and 124.18°. The maximum difference between the calculated contact angle of 119.44° and the estimated contact angles for invasion is 9.07°. For this report, we estimate $\theta_e$ as 119.44° (accurate to within 9.07°).

# 5 Supplementary Tables and Legends

Supplementary Table S1: Composition of dispersions used to create coatings

| Ingredient | NC1 Concentration (wt. %) | NC2 Concentration (wt. %) |
|---|---|---|
| PVDF | 0.5 | 0.5 |
| PMMA | 0.5 | 0.5 |
| HFS | 5.8 | 0.0 |
| PTFE | 0.0 | 5.8 |
| acetone | 88.9 | 88.9 |
| NMP | 4.3 | 4.3 |

Supplementary Table S2: Immersion experiments with gas-retaining insect surfaces by Balmert et al[6].
Surfaces in which microtrichia hair (sub-micron spacing) was present remained dry.

| Species (Location on insect) | Hair spacing (μm) Setae | Hair spacing (μm) Microtrichia | Liquid pressure | Dry/Wet, (duration) |
|---|---|---|---|---|
| Galerucella (Elytra) | 33 ± 14.8 | N/A | Ambient | Wet (2 days) |
| Corixa (Abd. Sternites) | 11 ± 2.3<br>103 ± 42.4 | N/A | Ambient | Wet (2 days) |
| Ilyocoris (Abd. Sternites) | 16 ± 4.4<br>147 ± 69.3 | N/A | Ambient | Wet (2 days) |
| Gerris (Sternites) | 8 ± 2.8 | 0.61 ± 0.2 | Ambient | Dry (11 days) |
| Ilyocoris (Elytra) | N/A | 0.4 ± 0.14 | Ambient | Dry (>120 days) |
| Notonecta (Elytra) | 81 ± 44.9 | 0.16 ± 0.07 | Ambient | Dry (>120 days) |

Abd. = Abdominal

# 6 Supplementary Figures and Legends

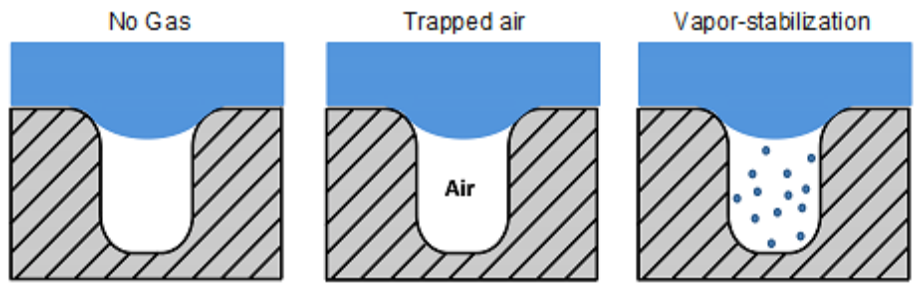

Supplementary Figure S1: Illustration of practically dry surfaces under water. Left: The roughness groove is not occupied by a gas. Middle: Air is trapped within the roughness groove. Right: Vapor fills the roughness groove and is in equilibrium with the overlying liquid.

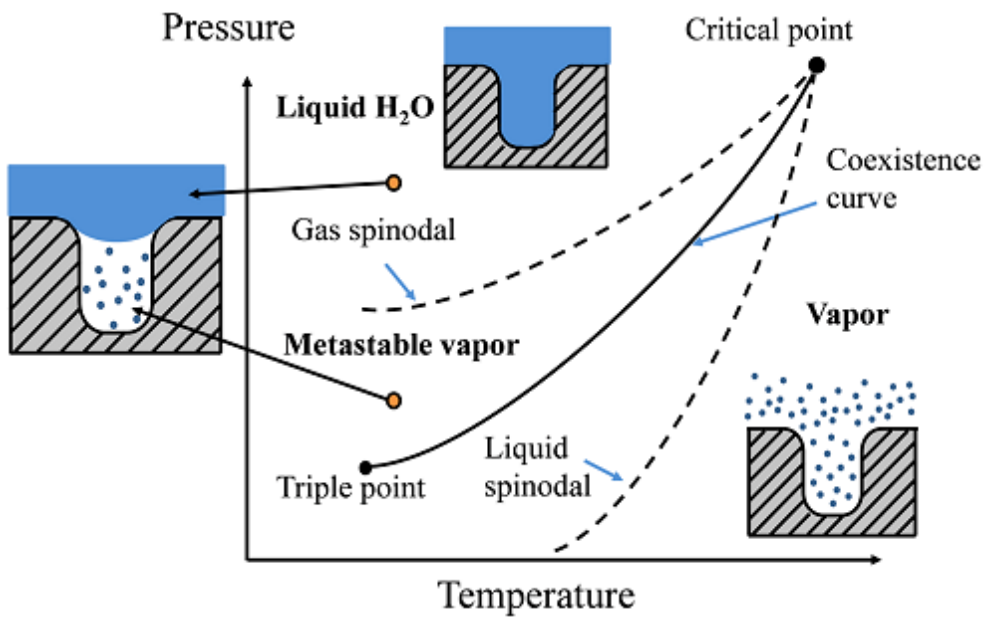

Supplementary Figure S2: Typical liquid-vapor phase diagram for water. The textured surface is superimposed on this diagram to indicate the phase of water within surface grooves. Here, we are designing for a metastable vapor to exist within surface grooves, and a stable liquid outside the grooves.[7,8]

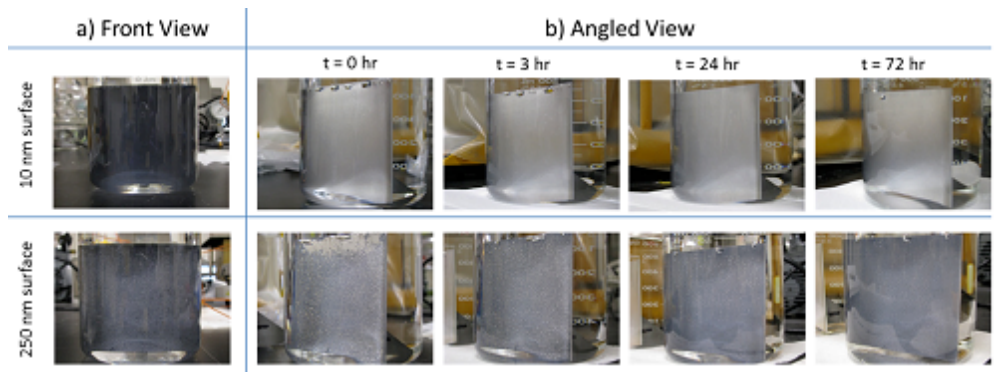

Supplementary Figure S3: Aging experiment. Aluminum substrate samples immersed underwater.

a) Front view of submerged surfaces. b) Angled view of submerged surfaces at various times after initial immersion.

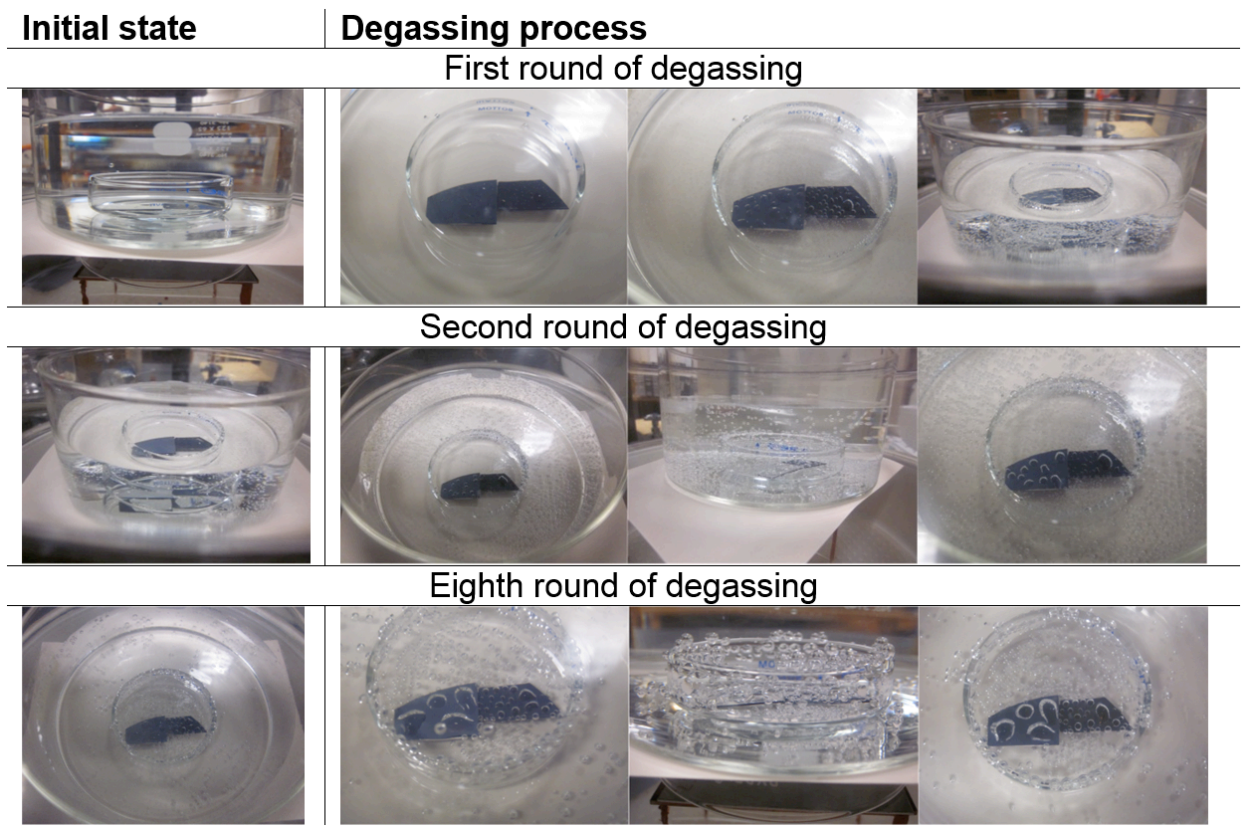

Supplementary Figure S4: Degassing process for zinc oxide coated sample in vacuum desiccator.

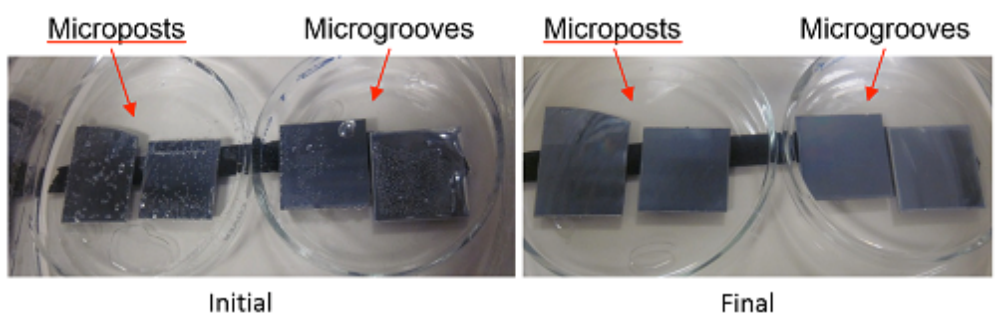

Supplementary Figure S5: Immersed silicon microposts and microgrooves samples. Samples are shown after three days of degassing in the vacuum desiccator.

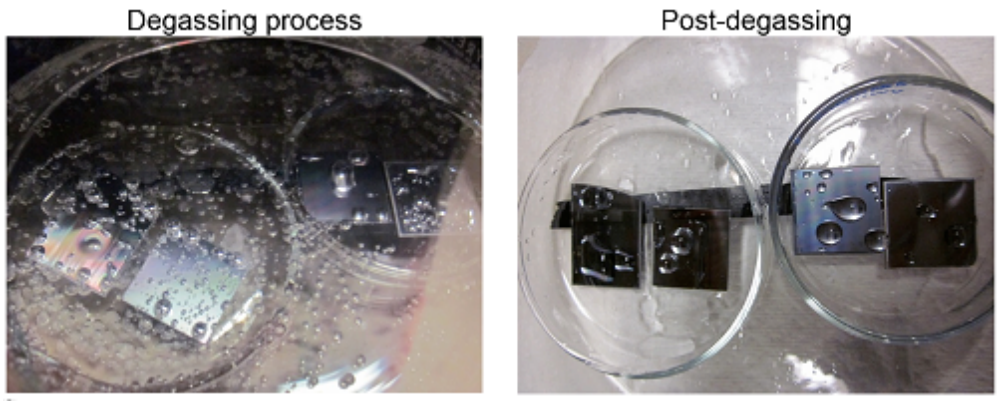

Supplementary Figure S6: Degassing process for silicon microposts and microgrooves samples. The samples are wet after degassing in the vacuum oven.

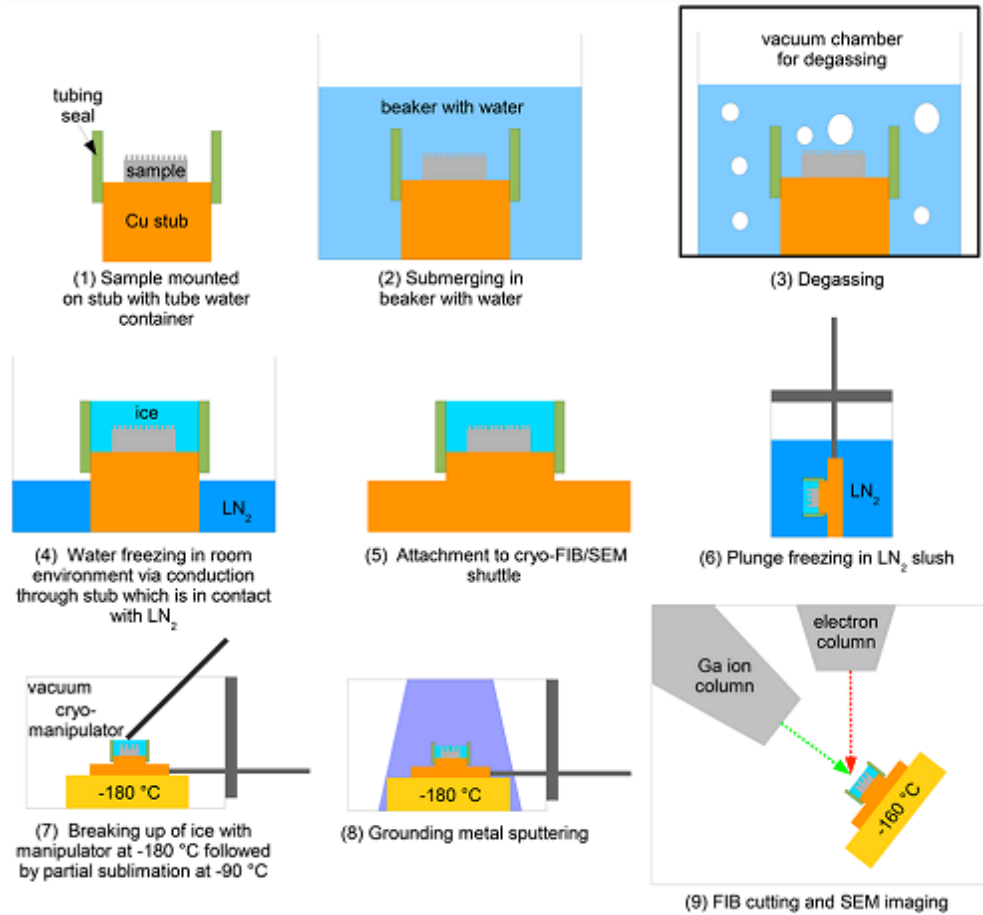

Supplementary Figure S7: Schematic of process employed for direct imaging of the liquid-solid interface.

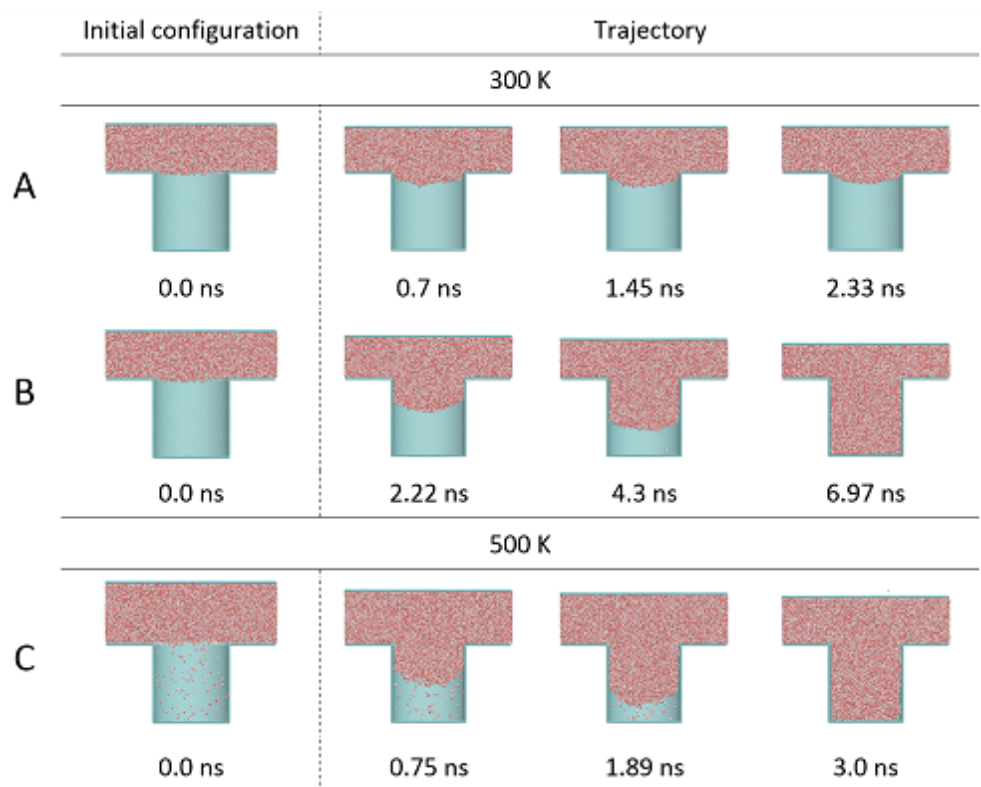

Supplementary Figure S8: Molecular dynamics simulations of a hydrophobic pore that is initially empty.

A) Hydrophobic pore demonstrating non-wetting at 300 K, 107.79 bar applied pressure. B) Hydrophobic pore that fully wets the surface at 300 K, 127.39 bar applied pressure. C) Hydrophobic pore that fully wets the surface at 500 K, 78.39 bar applied pressure.

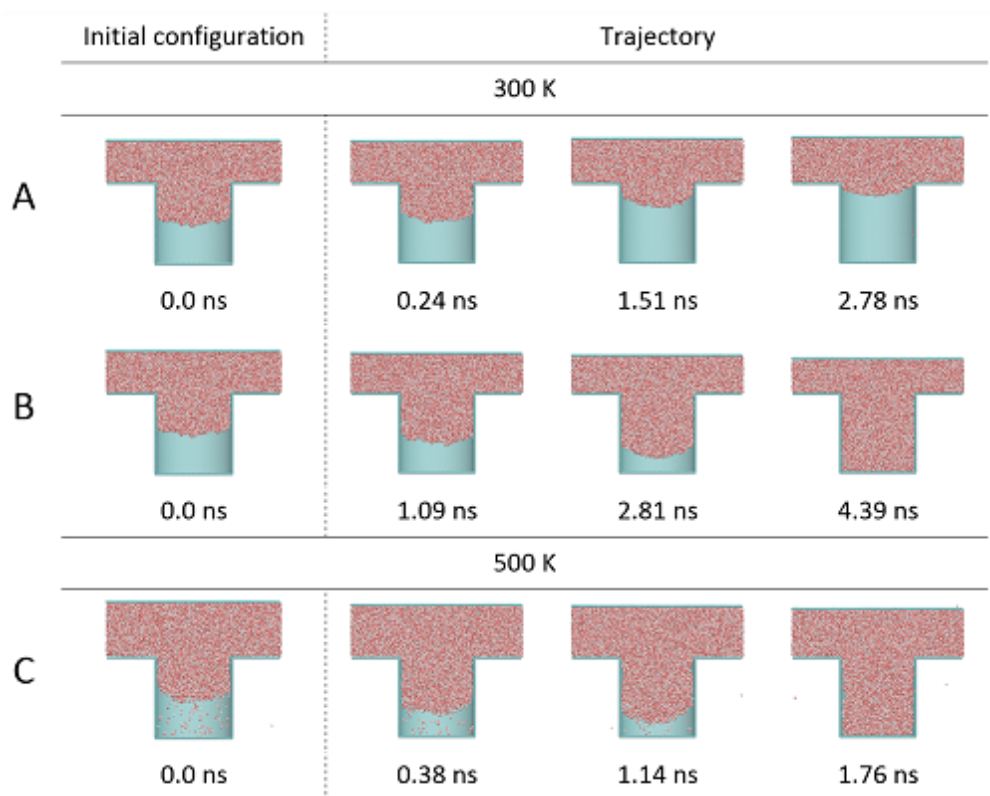

Supplementary Figure S9: Molecular dynamics simulations of a hydrophobic pore that is initially half-filled.

A) Hydrophobic pore demonstrating non-wetting at 300 K, 107.79 bar applied pressure. B) Hydrophobic pore that fully wets the surface at 300 K, 127.39 bar applied pressure. C) Hydrophobic pore that fully wets the surface at 500 K, 78.39 bar applied pressure.

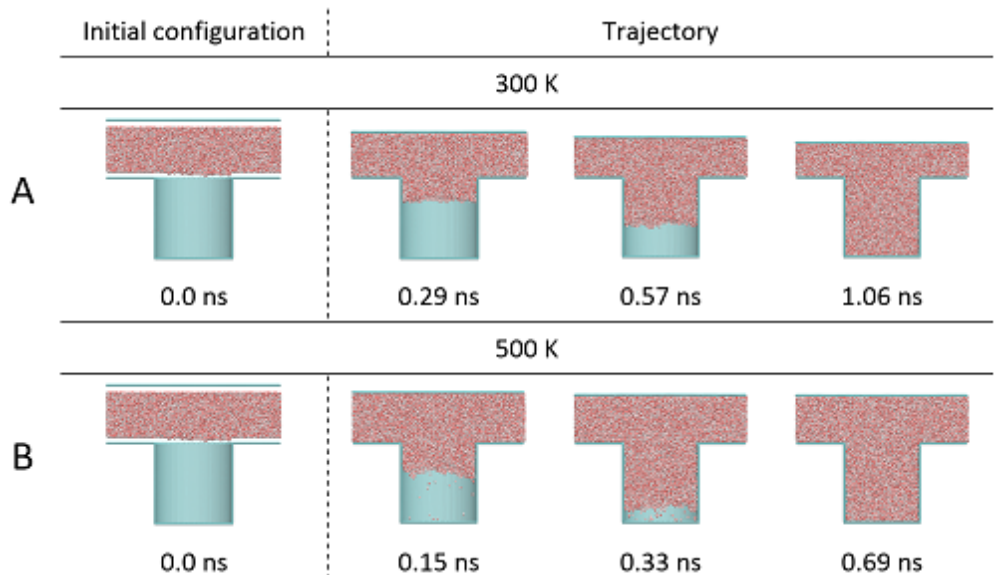

Supplementary Figure S10: Molecular dynamics simulations of a hydrophilic pore that is initially empty.

A) Hydrophilic pore that fully wets the surface at 300 K, 0 bar applied pressure. B) Hydrophilic pore that fully wets the surface at 500 K, 0 bar applied pressure.